# Experimental Demonstration of XOR Operation in Graphene Magnetologic Gates at Room Temperature

Hua Wen[1,2], Hanan Dery[3, †], Walid Amamou[1], Tiancong Zhu[2], Zhisheng Lin[1], Jing Shi[1], Igor Žutić[4], Ilya Krivorotov[5], Lu J. Sham[6, #], Roland K. Kawakami[1,2, ‡]

[1] Department of Physics and Astronomy, University of California, Riverside, Riverside, CA, 92521, USA

[2] Department of Physics, The Ohio State University, Columbus, OH, 43210, USA

[3] Department of Electrical and Computer Engineering, University of Rochester, Rochester, NY 14627, USA

[4] Department of Physics, University at Buffalo, State University of New York, Buffalo, NY, 14260 USA

[5] Department of Physics and Astronomy, University of California, Irvine, Irvine, CA, 92697, USA

[6] Department of Physics, University of California, San Diego, La Jolla, CA, 92697, USA

[‡]kawakami.15@osu.edu
[†]hdery@ur.rochester.edu
[#]lsham@ucsd.edu

**Abstract**

We report the experimental demonstration of a magnetologic gate built on graphene at room temperature. This magnetologic gate consists of three ferromagnetic electrodes contacting a single layer graphene spin channel and relies on spin injection and spin transport in the graphene. We utilize electrical bias tuning of spin injection to balance the inputs and achieve "exclusive or" (XOR) logic operation. Furthermore, simulation of the device performance shows that substantial improvement towards spintronic applications can be achieved by optimizing device parameters such as device dimensions. This advance holds promise as a basic building block for spin-based information processing.

Spintronics is an approach to electronics that utilizes the spin of the electron for information storage and processing [1-3]. By providing the ability to integrate logic with nonvolatile storage in ferromagnetic data registers, spintronics could greatly reduce the power consumption in logic circuits and go beyond traditional CMOS architectures. The demonstration of spin injection into semiconductors [4,5] prompted a variety of proposals for spintronic devices taking advantage of the tunable nature of semiconductors [6-11]. Among these was a proposal by Dery and Sham [12] for an "exclusive or" (XOR) gate based on spin accumulation in a semiconductor channel contacted by three ferromagnetic (FM) electrodes (see Fig. 1(a)). In this device, the magnetization directions of the first two FM electrodes represent the logic inputs ('0' and '1'), and spin injection from these input electrodes generates a current through the third FM electrode which represents the logic output. Subsequently, a more general proposal was developed that combines two such XOR gates to form a universal reconfigurable magnetologic gate (MLG) [8]. This MLG consists of five FM electrodes and the logic operation is represented by OR(XOR(A, B), XOR(C, D)), where A, B, C and D are the four logical input states and the fifth FM electrode reads the output. This can also be utilized as a universal two-input gate, where B and D define the gate operation (e.g. NAND, OR) and A and C represent the two inputs. The experimental discoveries of room temperature spin transport [13] and efficient spin injection into graphene [14] provided an ideal materials platform to realize such MLG devices. Motivated by these advances, the theoretical performance of graphene-based MLG was analyzed and novel spintronic circuits for rapid parallel searching were developed [15]. However, despite these extensive advances in the device modeling and spintronic circuit design, the experimental demonstration of the proposed three-terminal XOR and five-terminal universal MLG has been lacking.

In this Letter, we experimentally demonstrate the proposed three-terminal XOR magnetologic gate operation in a graphene spintronic device at room temperature. By carefully tuning the bias current between the two input electrodes, and an offset voltage in the detection loop, a clear non-zero output current (logic '1') is observed when the two inputs are antiparallel, with an absolute zero output current (logic '0') when the two inputs are parallel. These results provide the proof-of-concept demonstration for a class of magnetologic devices based on spin accumulation and establishes the feasibility of the universal five-terminal MLG. Furthermore, the signal size of the logic '1' output can be significantly enhanced by reducing the device size according to numerical simulation, making it promising for future spintronic applications.

The device geometry and measurement circuit are shown in Fig. 1(b). A flake of mechanically exfoliated single layer graphene is contacted by ferromagnetic cobalt (Co) electrodes A, B and M through MgO tunnel barriers [14,16]. The source current $I_S$ is a combination of $I_{AC}$ (ac current to inject spins) and $I_{DC}$ (dc bias current). The output voltage $V_{OUT}$ (ac voltage) is measured using standard low frequency lock-in techniques, and output current $I_{OUT}$ ( $\equiv V_{OUT}/R_{sen}$) is determined using a current detection scheme by systematically tuning the variable sensing resistor $R_{sen}$ [17]. An offset voltage $V_{OFFS}$ (ac voltage, phase and frequency locked to $I_{AC}$) is used to eliminate any background signal unrelated to spin. Reference electrode R (Ti/Au) is fabricated at the end of graphene and used as the ground point. Backgate voltage $V_G$ is applied on the Si substrate to tune the graphene carrier density.

Figure 1(c) shows the experimental demonstration of the XOR logic operation for a representative device. The four different input states are realized by sweeping an external magnetic field (H, collinear with the easy axes of the ferromagnets) to individually switch the magnetizations of input electrodes A and B, which have different magnetic shape anisotropy. The magnetization of M is kept downward during the logic operation. The measured $I_{OUT}$ varies with the different input states and demonstrates the XOR logic operation. When the inputs are parallel ('00' or '11'), $|I_{OUT}|$ is less than 0.023 nA. When the inputs are antiparallel ('01' or '10'), $|I_{OUT}|$ is stable at about 0.11 nA. The truth table of this XOR gate is summarized in the inset of Fig. 1(c). In the rest of the paper, we explain how this XOR logic operation is achieved and how the output signal could be optimized for future applications.

An important preliminary step is validation of the spin transport properties using traditional nonlocal voltage detection [13,18]. For the measurement circuit in Fig. 1(b), bias current $I_{DC}$ and offset voltage $V_{OFFS}$ are set to zero and $R_{sen}$ is adjusted to be sufficiently large (10 MΩ) to perform voltage detection [17]. For the device under investigation, the Dirac point is located at $V_G = -13$ V [19], and $V_G$ is set to +30 V for the measurements. Electron spins are injected through inputs A and B using a current of $I_{AC}$ = 1 µA (11 Hz). Figure 2(a) shows the nonlocal voltage $V_{OUT}$ at different magnetization states of A, B and M. We observe three jumps in $V_{OUT}$ as the magnetic field H is swept upward or downward, which correspond to the magnetization switching of the three ferromagnets. This indicates successful spin injection, transport and detection in our device [20].

In order to examine the logic operation of our device, the output electrode needs to be maintained at fixed magnetization. This is possible because electrode M has a distinct, and, more importantly, larger coercive field than A and B. It is worth noting that the coercive field of the electrodes are different for positive and negative fields due to the geometrical shape of the electrodes, which can create domain wall pinning [16,21]. However, it is found that the magnetic field required to switch M from ↓ to ↑ is significantly larger than that of A and B, as shown in Fig. 2(a). Therefore, M is initially magnetized to be ↓ and then H is swept below +44 mT. In this way, the magnetization state of M is fixed at ↓ throughout the logic operation.

Figure 2(b) shows the voltage signal $V_{OUT}$ with four different input states (↓↓, ↑↓, ↑↑ and ↓↑) when H is swept between -20 mT and +40 mT. The input states are realized in the following order: ↓↓, ↑↓, ↑↑, ↓↑, ↓↓ when H is swept through -20 mT → +40 mT → -20 mT. We observe that for antiparallel inputs, $V_{OUT}$ has different values compared to parallel inputs. However, two challenges need to be overcome in order for $V_{OUT}$ to produce the proper logic output signal. The first challenge is that input A contributes a smaller signal to the output compared to input B due to the fact that input A is further away from M than input B. This is indicated by the different values of $\Delta V_A$ and $\Delta V_B$ in Fig. 2(b). The second challenge is that $V_{OUT}$ is not zero for parallel inputs. These two problems make it difficult to discriminate between logic '0' and '1'. In the following, we present our methods to resolve these challenges.

To tune the signal contribution of inputs A and B, a DC bias current $I_{DC}$ is added in the injection current loop as shown in Fig. 1(b), where positive $I_{DC}$ is defined as current flowing from B, through the graphene, to A. It was previously shown that the nonlocal spin signal can be significantly tuned by a DC bias current [22,23]. Similar bias dependence is observed in our devices, with the spin signal increasing at positive bias (current flowing from electrode to graphene) and decreasing at negative bias [19]. Because inputs A and B are under opposite bias when $I_{DC}$ flows through the injection circuit, we can tune spin signal from input A ($\Delta V_A$) and input B ($\Delta V_B$) in an opposite manner. This is illustrated in Fig. 3(a) and Fig. 3(b), where $I_{DC}$ is varied from -15 µA to +15 µA. As $I_{DC}$ is increased, the value of $\Delta V_A$ decreases while the value of $\Delta V_B$ increases. Notably, when $I_{DC}$ is at -7 µA, the spin signals from inputs A and B are equal ($\Delta V_A = \Delta V_B$). This results in the balanced output curve as shown in Fig. 3(a) for $I_{DC}$ = -7 µA. We have reproduced this tuning and balancing of the two inputs on multiple devices with MgO and $Al_2O_3$ tunnel barriers [19].

To tune the logic '0' output to actual zero, $V_{OFFS}$ is added in the detection loop (Fig. 1(b)). Output '0' level (defined as background signal $V_{bg}$) is systematically adjusted by varying $V_{OFFS}$ [19]. For the device presented in this paper, $V_{bg}$ is close to zero when $V_{OFFS}$ = +8 µV.

In order to utilize this spin-based MLG in the proposed spintronic circuit [8,15], we need to convert the logic output from a voltage signal to a current signal. The current output allows the integration of multiple XOR MLGs before doing spin-to-charge conversion. This can greatly reduce the power consumption when performing large data search applications by using this XOR gate as compared to traditional CMOS devices [15]. By utilizing the current detection scheme developed for graphene spin valves [17], we achieve a current output for our XOR logic with sufficiently small $R_{sen}$ (1-3 kΩ, see [19]). The resulting current output signal is shown in Fig. 1(c), with $R_{sen}$ = 3 kΩ, $I_{DC}$ = -7 µA and $V_{OFFS}$ = +8 µV. This curve displays precisely the behavior needed for the XOR magnetologic gate [12]. When the two inputs are '00' or '11', the output current is zero (logical '0'), and when the two inputs are '01' or '10', the output current is nonzero (logical '1'). We note that the output currents of '01' and '10' logic states have opposite polarities. On one hand, this can be rectified in circuits where the opposite polarity is undesirable. On the other hand, when this XOR gate is considered as a building block for the more complex five-terminal MLG, the opposite polarity of the output is essential [8,15].

While this proof-of-concept demonstration of a graphene-based magnetologic gate shows promise for future spintronics devices, two key improvements are needed for practical applications of the gate.

First, writing of the magnetic information should be facilitated by spin-transfer torque (STT) techniques [24] or spin Hall effect [25,26] (SHE) that alleviates the need for external magnetic fields and different shapes for contacts. We note that STT and SHE can be achieved by an all metallic path, with no current leakage to graphene [14]. Second, the operation should be independent of device-specific bias current ($I_{DC}$) and offset voltage ($V_{OFFS}$). For the former, this can be achieved by engineering device parameters including spin polarization of the contacts, geometric size of contacts and graphene, the spin diffusion length of graphene, etc., and is further discussed in the following paragraphs. For the latter, $V_{OFFS}$ is used to cancel the *spin-independent* signal that is present in many nonlocal spin valve measurements, but whose origin is unknown. However, in spintronic circuits of the type proposed by Dery and Sham [8,15], the oscillating magnetization of the readout electrode (M) will extract only the *spin-dependent* part of the signal, thereby alleviating the need for $V_{OFFS}$.

We simulate the output current signal for various critical device parameters using one dimensional spin drift-diffusion model considering finite-size contacts [19,27]. The current spin polarization of electrode A, B and M are assumed to be of the same ($P_J$). $P_J$ is experimentally measured to be ~0.11 in the presented device. Figure 4(a) shows the signal difference ($\Delta I_{OUT}$) between '1' and '0' output ($\Delta I_{OUT} \equiv |I_{OUT}('1') - I_{OUT}('0')|$) for $P_J$ = 0.11, 0.20, 0.30. Increasing $P_J$ is found to significantly increase $\Delta I_{OUT}$. $\Delta I_{OUT}$ can be enhanced even further with higher $P_J$ using alternative tunnel barriers [28]. Interestingly, we find that contact resistance of M ($R_M$) plays an important role in optimizing the output current. $I_{OUT}$ exhibits a maximum at $R_M$ ~ 1.2 k$\Omega$. This optimal $R_M$ depends on the graphene size and the spin diffusion length of graphene (2.2 µm in the present device as determined through Hanle precession [20]). For lower $R_M$, contact induced spin relaxation reduces the spin accumulation and thus reduces the output current [29]. For higher $R_M$, the increased resistance of the detection circuit reduces the output current [17]. The current at optimal $R_M$ is about 2 times larger than the output current for $R_M$ ~ 11.5 k$\Omega$ in the presented device (grey dot in Fig. 4(a)).

The performance of the device can be further improved by working in a confined geometry. Figure 4(b) shows the simulated current for logic '0' and '1' output for a much smaller device (shown in the inset). This device has a size of 350 nm × 500 nm (L × W) and there is no graphene extending outside of electrodes A and R [19]. The current for '1' output is two orders of magnitude larger than '0' output due to the reduced spacing (center-to-center distance is 100 nm) between the two inputs, A and B. In such a confined device, the logic operation is simplified since there is no need for adding $I_{DC}$ to balance the contribution of inputs A and B because the electrode spacing is much less than the spin diffusion length [30]. This simplification is crucial for large-scale integration of these devices into logic circuits. Whereas scaling down the feature size in modern CMOS technology leads to undesirable leakage currents, the performance of our MLGs will continue to improve with further reduction of the distance between contacts [15].

In conclusion, we have demonstrated a graphene MLG that performs XOR logic at room temperature. The key step is to systematically tune the injection current bias to balance the contributions of the two input ferromagnetic electrodes to the output signal. With further reduction of the graphene area and optimization of the magnetic contacts (resistance and spin polarization), these MLGs will improve the performance of information-processing integrated circuits.

**Acknowledgements**

The authors acknowledge Y. Kato and B. Bushong for helpful discussions. This work was supported by NRI-NSF (DMR-1124601). We also acknowledge the support from ONR (N00014-14-1-0350), NSF (DMR-1310661), NSF-MRSEC (DMR-1420451), NSF (ECCS-1231570), NSF (ECCS-1508873), DTRA (HDTRA1-13-1-0013), ONR (N00014-13-1-0754) and C-SPIN, one of the six SRC STARnet Centers, sponsored by MARCO and DARPA.

FIG. 1. Experimental demonstration of graphene XOR magnetologic gate. (a) Diagram of proposed XOR magnetologic gate device. A, B and M are ferromagnetic electrodes on top of a spin transport channel. Input logic '1' and '0' are the two magnetization directions along the easy axis of the electrodes. $I_S$ injects spins through the two inputs, A and B. $I_{OUT}$ is the logic output signal. (b) Cartoon of experimental device structure and measurement setup. A, B and M are MgO/Co electrodes. Spin channel is a single layer graphene. R is Ti/Au nonmagnetic reference electrode used as ground point. $I_{OUT}$ and $V_{OUT}$ are the measured current and voltage signal, respectively. $R_{sen}$ is a variable resistor. $V_{OFFS}$ is an ac voltage source. External magnetic field H is applied to the easy axis of the electrodes. Center-to-center distance of the electrodes are: $L_{AB}$ = 1.6 μm, $L_{BM}$ = 1.8 μm. $L_{MR}$ = 7.85 μm. Graphene width along H direction is ~4.3 μm. (c) $I_{OUT}$ measured as a function of H. Black (red) curve indicates H sweeps upwards (downwards). Vertical arrows indicate the magnetization states of A and B. Top left inset: truth table of XOR logic operation.

FIG. 2. Nonlocal voltage detection of spin transport. (a) and (b) Voltage signal $V_{OUT}$ as a function of H for a full sweep ((a), -45 mT to +60 mT) and minor loop ((b), -20 mT to +40 mT). $I_S = I_{AC}$ = 1 μA. In (b), only A and B switch their magnetization. The change of $V_{OUT}$ when A (B) switches its magnetization direction is noted as $\Delta V_A$ ($\Delta V_B$). M is fixed to be ↓.

FIG. 3 Tuning $\Delta V_A$ and $\Delta V_B$ using bias current $I_{DC}$. (a) $V_{OUT}$ as a function of H (minor loop, as in Fig. 2(b)) at different $I_{DC}$. Curves are shifted vertically for clarity. (b) $\Delta V_A$ and $\Delta V_B$ as a function of $I_{DC}$. At positive (negative) $I_{DC}$, B is under positive (negative) bias and A is under negative (positive) bias. Vertical arrows indicates the flow of current $I_{DC}$. 'Gr.' represents graphene channel.

FIG. 4. Optimizing output current signal. (a) Signal difference between '1' and '0' logic output $\Delta I_{OUT}$ ($\equiv |I_{OUT}('1') - I_{OUT}('0')|$) as a function of $R_M$ for different spin polarization of contacts, assuming $P_A = P_B = P_M = P_J$, and $P_R = 0$. Grey dot represents our current device parameters. (b) Output signal $I_{OUT}$ ('1' and '0') as a function of $R_M$ for an optimized device geometry. Signal for '0' is magnified by 10 times. Inset: optimized device structure. There is no graphene beyond electrode A and R. The whole device length is L = 350 nm. Graphene width is W = 500 nm. Each electrode (A, B, M and R) has width of 50 nm, and center-to-center distance between adjacent electrodes is 100 nm. Spin polarization $P_J$ is 0.30.

Figure 1:

Figure 2:

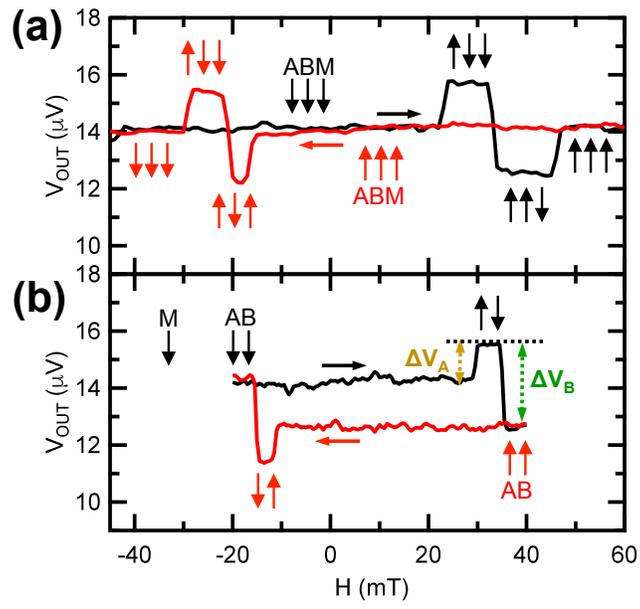

Figure 3:

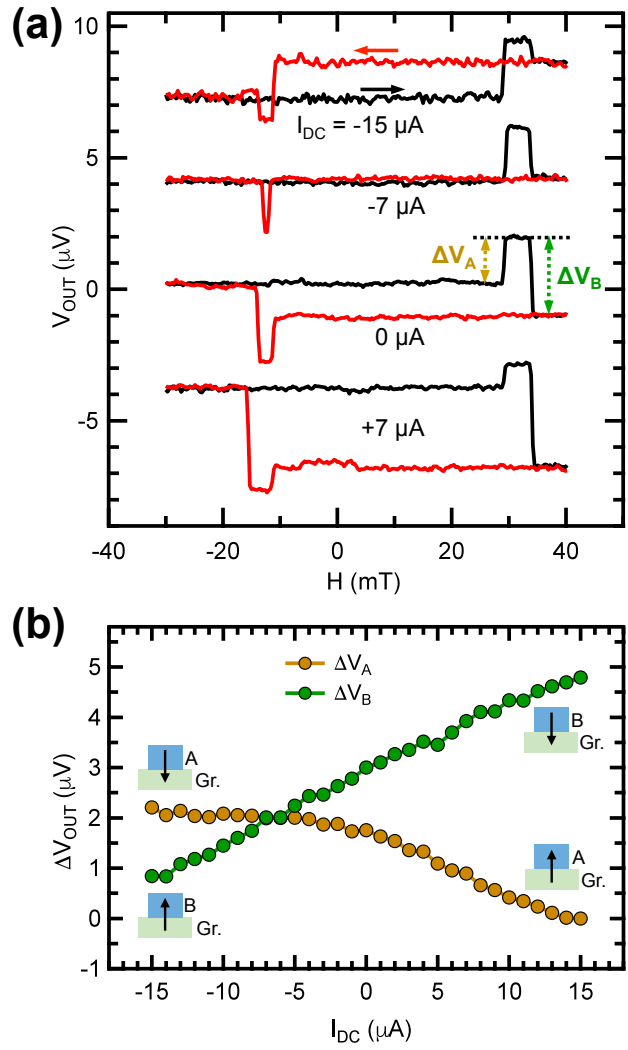

Figure 4:

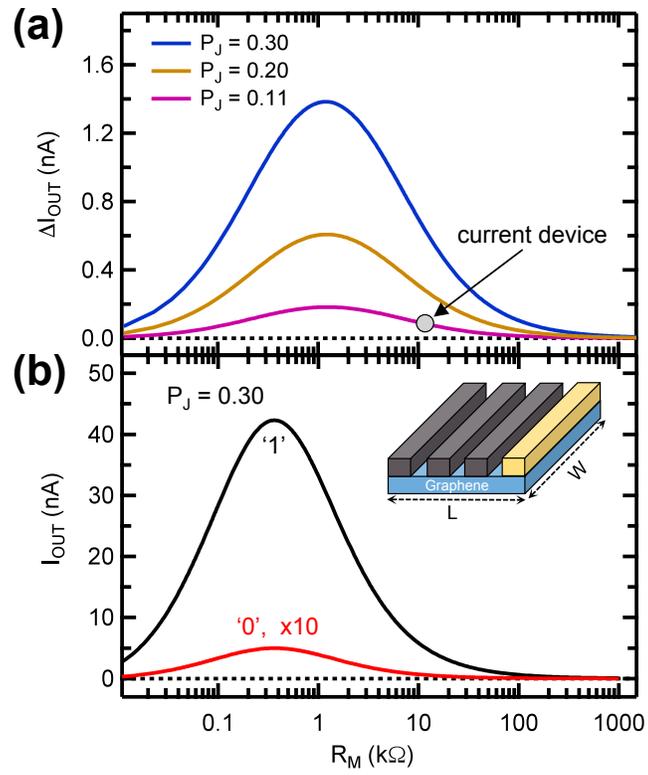

# Supplemental Material:

**Experimental Demonstration of XOR Operation in Graphene Magnetologic Gates at Room Temperature**

Hua Wen[1,2], Hanan Dery[3,†], Walid Amamou[1], Tiancong Zhu[2], Zhisheng Lin[1], Jing Shi[1], Igor Žutić[4], Ilya Krivorotov[5], Lu J. Sham[6,#], Roland K. Kawakami[1,2,‡]

[1] Department of Physics and Astronomy, University of California, Riverside, Riverside, CA, 92521, USA

[2] Department of Physics, The Ohio State University, Columbus, OH, 43210, USA

[3] Department of Electrical and Computer Engineering, University of Rochester, Rochester, NY 14627, USA

[4] Department of Physics, University at Buffalo, State University of New York, Buffalo, NY, 14260 USA

[5] Department of Physics and Astronomy, University of California, Irvine, Irvine, CA, 92697, USA

[6] Department of Physics, University of California, San Diego, La Jolla, CA, 92697, USA

‡kawakami.15@osu.edu

†hdery@ur.rochester.edu

#lsham@ucsd.edu

**Section 1: Device SEM Image and Graphene Resistance Measurement**

Figure S1(a) shows the SEM image of the device presented in the main text of the paper. The graphene flake is outlined by the yellow dashed line. Two Ti/Au electrodes labeled "L" and "R" are fabricated on the two ends of the graphene and the other electrodes are Co with 1.2 nm MgO tunnel barriers. The 90° bend in the Co electrodes are used to pin domain walls to promote magnetization switching at different fields for the different electrodes. Three of the Co electrodes are chosen as A, B, and M. Figure S1(b) shows the graphene resistance for channel AM and BM as a function of backgate voltage $V_G$. The measurement utilizes a four probe geometry (shown in the top right inset), where electrical current (I) flows between L and R, and voltage V is measured between A and M (or between B and M). The graphene channel lengths of AM and BM are 3.4 μm and 1.8 μm, respectively, and the graphene width is 4.3 μm. The graphene shows bipolar transport with the charge neutrality point at $V_G = -13$ V and field effect mobility ~ 1700 cm$^2$/Vs which is typical for graphene devices on SiO$_2$ substrate.

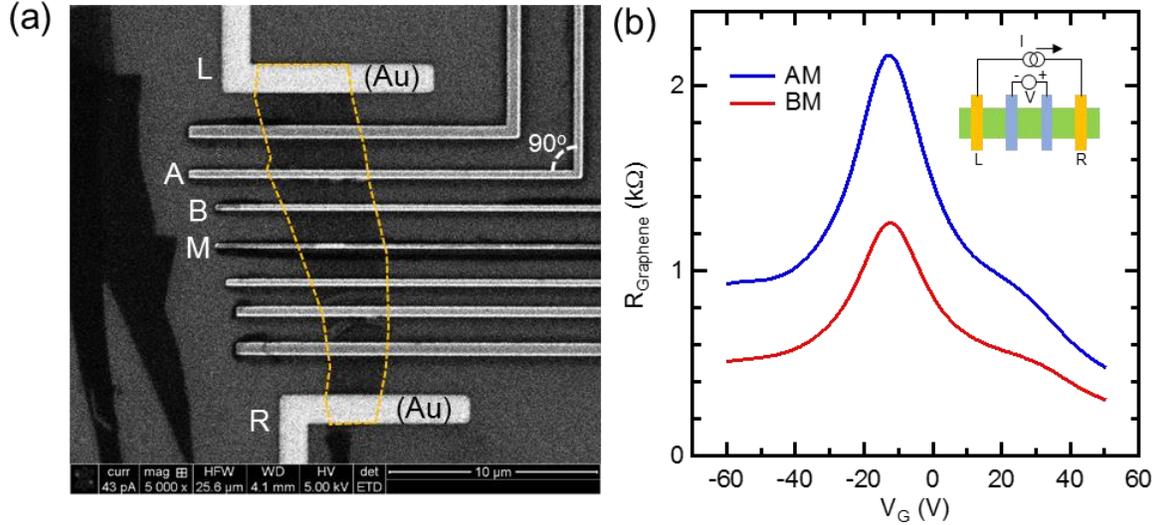

**Fig. S1.** (a) SEM image of the device presented in the main text. (b) Four-probe graphene resistance vs. $V_G$ for channel AM and BM.

**Section 2: Bias Dependence of Spin Injection**

Bias tuning of spin injection is employed in our experiment to obtain equal contributions from the two inputs. To determine the bias dependence of a single Co electrode, spin transport measurements are performed in the nonlocal geometry with nonmagnetic Au outer electrodes (inset of Fig. S2(a)). The nonlocal voltage $V_{NL}$ is measured by lock-in detection in response to an AC current excitation across the Co spin injector ($I_{AC} = 1$ μA). Figure S2(a) shows $V_{NL}$ as a function of magnetic field (H) for three different DC bias currents: $I_{DC} = -5.0$ μA, $-2.6$ μA, $+0.1$ μA. The black (red) curve is for the sweep with increasing (decreasing) H, and the vertical arrows indicate the magnetization states of the two Co electrodes. The spin signal $\Delta V_{NL}$ is defined as the nonlocal voltage difference between parallel and antiparallel magnetization states. For these measurements, the backgate voltage $V_G$ is set to the charge neutrality point ($V_{CNP}$) of $-2$ V. Comparing the three curves shows that the spin signal $\Delta V_{NL}$ changes sign with DC bias current. A more detailed dependence of $\Delta V_{NL}$ on DC bias current is shown in Fig. S2(b), where the scan is repeated for three different backgate voltages: $V_G = V_{CNP}$, $V_{CNP}$ -30 V, and $V_{CNP}$ +30 V. For all three cases, the bias dependence shows a general increase of $\Delta V_{NL}$ with $I_{DC}$. We have observed this type of bias dependence on numerous devices. The physical origin of this bias dependence could be from the ferromagnetic density of states [S1, S2, S3], surface states at the interface [S4], or other effects and is still under investigation.

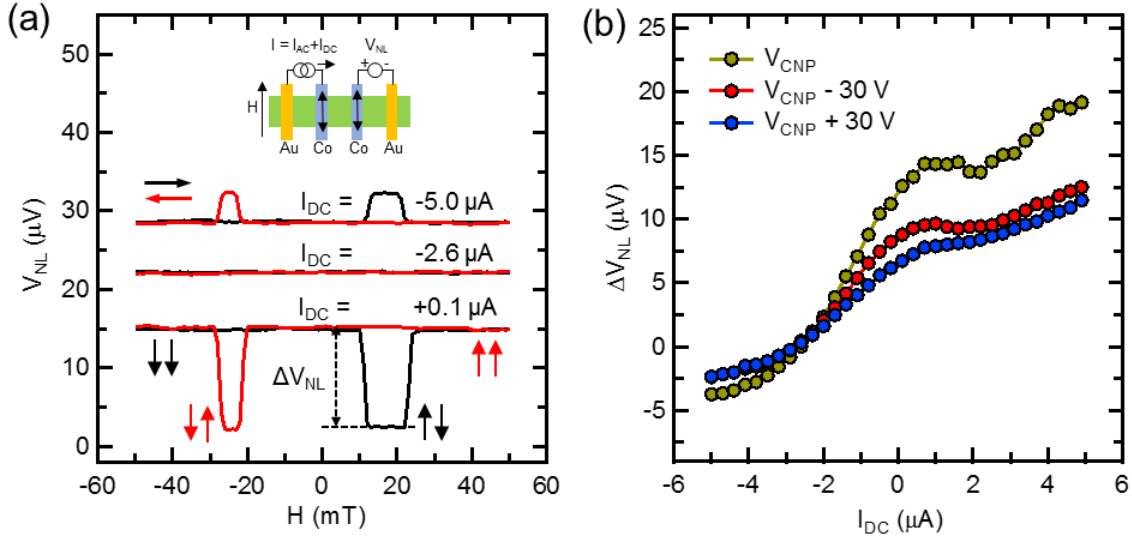

**Fig. S2.** Typical bias dependence of spin injection. (a) Measured nonlocal voltage $V_{NL}$ as a function of magnetic field H for three different DC bias currents. Inset: diagram of the nonlocal measurement. (b) $\Delta V_{NL}$ as a function of $I_{DC}$ for three different backgate voltages.

**Section 3: XOR Operation using Al$_2$O$_3$ Tunnel Barrier**

Al$_2$O$_3$ has been widely used in spintronic devices such as magnetic tunnel junctions as a stable, uniform and easy-to-fabricate tunnel barrier. In order to show the versatility of our XOR device, we fabricate graphene spin valves with uniform Al$_2$O$_3$ tunnel barriers. The Al$_2$O$_3$ barrier is achieved by sputtering ~0.6 nm Al on top of exfoliated single layer graphene flakes followed by oxidization in ~750 Torr O$_2$ for 30 min to convert Al to Al$_2$O$_3$ [S5]. Standard e-beam lithography is then carried out to define

Co electrodes using MMA/PMMA bilayer resist. 80 nm of Co is deposited on top of $Al_2O_3$ in a molecular beam epitaxy system (base pressure $\sim 1\times10^{-10}$ Torr). Finally, the device is lifted off in PG Remover.

We demonstrate XOR logic on these $Al_2O_3$ tunnel barrier devices (Fig. S3) using the same measurement procedures described in the main text. About 0.06 nA current is observed for the '1' output. During the logic operation, the magnetization of M is maintained at ↓. The other parameters used in the measurement are: $I_{AC} = 1$ µA, $I_{DC} = 3$ µA, $V_{OFFS} = 6$ µV, $R_{sen} = 15$ kΩ, $V_G = +20$ V. The success of $Al_2O_3$ tunnel barrier shows that our XOR devices are compatible with large volume fabrication for application purposes.

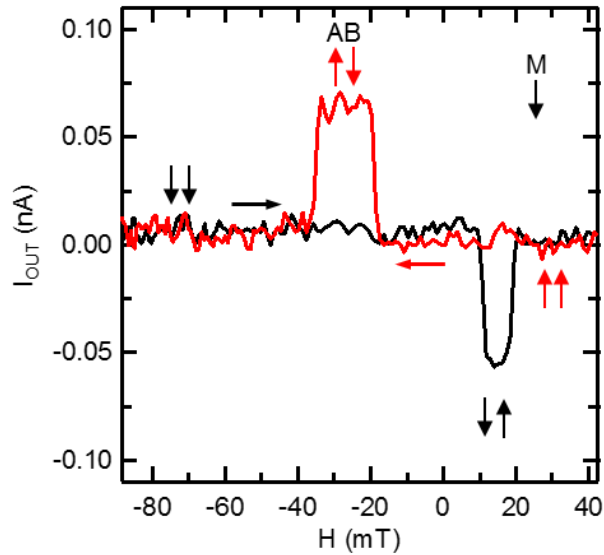

**Fig. S3** XOR logic operation for graphene magnetologic devices with $Al_2O_3$ tunnel barrier. $I_{OUT}$ as a function of magnetic field H. Horizontal black (red) arrow indicates up (down) sweep of H. Vertical arrows indicate the magnetization configuration of A, B and M.

### Section 4: Output Signal Tuning using $V_{OFFS}$

Figure S4(a) shows $V_{OUT}$ as a function of magnetic field (H) for different values of $V_{OFFS}$ at the optimal bias condition ($I_{DC} = -7$ µA) for the device presented in the main text. First, we observe that the measured background voltage $V_{bg}$ can be tuned systematically with $V_{OFFS}$. Second, we notice that the voltage jumps at different $V_{OFFS}$ have the same magnitude. This implies that adding $V_{OFFS}$ does not corrupt the optimal bias condition for matching the two inputs. Figure S4(b) shows that $V_{bg}$ changes linearly with $V_{OFFS}$. As a result, we can tune $V_{bg}$ to zero by setting $V_{OFFS} = 8$ μV.

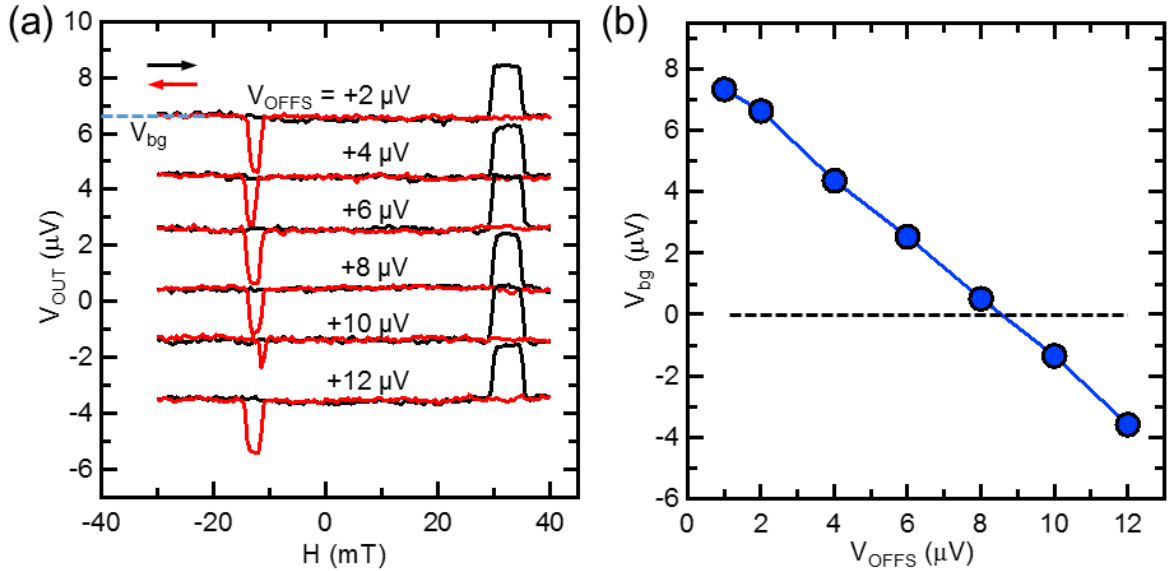

**Fig. S4** Tuning background of $V_{OUT}$ using offset voltage $V_{OFFS}$. (a) Minor loop of $V_{OUT}$ at different offset voltage $V_{OFFS}$. $V_{OUT}$ at logic '0' is noted as $V_{bg}$. (b) $V_{bg}$ as a function of $V_{OFFS}$.

## Section 5: $I_{OUT}$ at Different $R_{sen}$

As described in [S6], when $R_{sen} \rightarrow 0$, $I_{OUT}$ approaches a constant value, which indicates current-based detection of spin accumulation. For the device presented in the main text, we observe little change of $I_{OUT}$ when $R_{sen}$ is changed from 3 kΩ to 1 kΩ (Fig. S5). This means we are in the current detection regime.

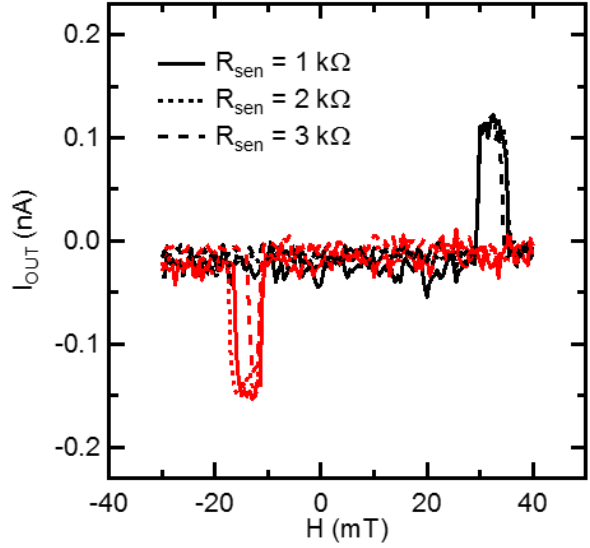

**Fig. S5** $I_{OUT}$ as a function of H for three different $R_{sen}$ values: 1 kΩ, 2 kΩ and 3 kΩ for the device presented in the main text. Horizontal black (red) arrow indicates up (down) sweep of H. Vertical arrows indicate the magnetization configuration of the two inputs, A and B.

**Section 6: Simulation of $I_{OUT}$ using Drift-Diffusion Model**

In the one dimensional spin drift-diffusion model [S7], the effect of the contacts on the current distribution is taken into account. This is critical when contact size is comparable to device dimension. To simulate $I_{OUT}$ at different $P_J$ (Fig. 4(a)) for the device presented in the main text, the following parameters are used: graphene width $W_G$ = 4.3 μm, graphene sheet resistance $R_\square$ = 1.2 kΩ at $V_G$ = +30 V, spin diffusion length of graphene $\lambda_G$ = 2.2 μm as acquired from Hanle precession measurement. The spin injection current is $I_{AC}$ = 1 μA. Center-to-center distance of the electrodes are: $L_{AB}$ = 1.6 μm, $L_{BM}$ = 1.8 μm, $L_{MR}$ = 7.85 μm. Width of the electrodes are $W_A$ = 0.377 μm, $W_B$ = 0.272 μm, $W_M$ = 0.194 μm, $W_R$ = 1.5 μm. There is 4.6 μm of graphene extending outside A, and no graphene extending outside of R. The variable resistor $R_{sen}$ = 3 kΩ. Contact resistances of the electrodes are $R_A$ = 10.9 kΩ, $R_B$ = 4.8 kΩ, $R_R$ = 0.2 kΩ. $R_M$ is varied in the simulation. Antiparallel input is ↑↓, and parallel input is ↓↓.

To simulate $I_{OUT}$ for a device in the confined geometry (Fig. 4b), the following parameters are used: $W_G$ = 500 nm, $R_\square$ = 1.2 kΩ, $I_{AC}$ = 1 μA, $L_{AB} = L_{BM} = L_{MR}$ = 100 nm, $W_A = W_B = W_M = W_R$ = 50 nm, $P_J$ = 0.30, $R_{sen}$ = 3 kΩ, $R_A$ = 10.9 kΩ, $R_B$ = 4.8 kΩ, $R_R$ = 0.2 kΩ. $R_M$ is varied in the simulation. Antiparallel input is ↑↓, and parallel input is ↓↓.

**Supplemental Material Reference:**